# Signal processing in the TGF-β superfamily ligand-receptor network

Jose M. G. Vilar[*], Ronald Jansen, and Chris Sander

*Computational Biology Center, Memorial Sloan-Kettering Cancer Center, 307 E. 63rd St., New York, NY 10021*

## Abstract

**The TGF-β pathway plays a central role in tissue homeostasis and morphogenesis. It transduces a variety of extracellular signals into intracellular transcriptional responses that control a plethora of cellular processes, including cell growth, apoptosis, and differentiation. We use computational modeling to show that coupling of signaling with receptor trafficking results in a highly versatile signal-processing unit, able to sense by itself absolute levels of ligand, temporal changes in ligand concentration, and ratios of multiple ligands. This coupling controls whether the response of the receptor module is transient or permanent and whether or not different signaling channels behave independently of each other. Our computational approach unifies seemingly disparate experimental observations and suggests specific changes in receptor trafficking patterns that can lead to phenotypes that favor tumor progression.**

## Introduction

The TGF-β signal transduction pathway follows an apparently straightforward downstream cascade, progressing sequentially from the interaction of ligands with transmembrane receptors, through phosphorylation of mediator Smad proteins, to transcriptional responses (Figure 1). The simple logic of this signal transduction cascade strongly contrasts with the molecular complexity of the cellular processes involved and the wide diversity of responses triggered.

At the molecular level, there is an intricate signal transduction machinery that integrates signals from the 42 known ligands of the TGF-β superfamily, funnels them through the two principal regulatory Smad (R-Smad) channels (Smad1/5/8 or Smad2/3) and subsequently leads to the wide-spread transcriptional control of over 300 target genes in a cell-context dependent manner (Kang et al., 2003b) (see Figure 2). The components of this machinery include the members of the two main receptor families (type I and type II receptors), a myriad of adaptor proteins, and the trafficking apparatus of the cell, which shuttles proteins between different subcellular compartments. Each ligand induces the formation of a receptor complex with type I and type II receptors, which then signal through one of the two Smad channels (Goumans et al., 2003; Yamashita et al., 1994).

---

[*] Correspondence: vilar@cbio.mskcc.org

The ability of most ligands to bind several type I and type II receptors results in a complex ligand-receptor interaction network (Figure 2).

At the phenotypic level, the responses are extremely diverse. The members of the TGF-β superfamily act prototypically as potent negative growth regulators but, depending on the cell type and context, they can also induce differentiation, apoptosis, cell migration, adhesion and extracellular matrix deposition. TGF-β itself is of particular interest in cancer research. In epithelial cells, it suppresses cellular growth by inducing G1 arrest (mediated by transcriptional activation of p15 and p21) (Siegel and Massagué, 2003) and its inactivation contributes to tumorigenesis. The versatility of the pathway in eliciting different types of behavior is perhaps best epitomized by the pervasive, rather paradoxical ability of TGF-β to change its function from suppressor to promoter of growth in epithelial cells during tumor progression (Cui et al., 1996; Siegel and Massagué, 2003).

Current theories for explaining the variety of responses to members of TGF-β superfamily of ligands focus mainly on the downstream transcriptional regulatory networks they activate: transcriptional co-factors of the R-Smads are expressed at different levels in a cell-specific manner, thereby modifying downstream responses. In fact, the role reversal of TGF-β from negative to positive growth regulator has been found to be associated with a phenotypic change known as epithelial-to-mesenchymal transition, in which cells change the co-factors recruited by the R-Smads and acquire motile phenotypes (Cui et al., 1996; Gotzmann et al., 2004; Thiery, 2003).

It is striking, however, that such a variety of complex responses and intricate molecular components are connected through just two Smad channels by such a simple downstream signal transduction cascade. There is a richness of experimental observations that are difficult to reconcile with this observation. In particular, whether TGF-β acts as a growth suppressor or promoter can depend on whether the tumor cells were grown *in vitro* or *in vivo* (Steiner and Barrack, 1992). In these two different situations, the extracellular context determines the way in which cells respond to TGF-β. It has been suggested that TGF-β can suppress the growth of cells around the tumor, that it can shut down locally the immune system, and that it can promote angiogenesis. All these paracrine effects would help the growth of the tumor *in vivo*, where it has to compete with neighboring cells. So far, although appealing, none of these mechanisms has been identified as an alternative cause of the TGF-β role reversal.

The most direct way in which the extracellular context can affect the functioning of the TGF-β pathway is through signaling of other ligands of the TGF-β superfamily. As we have mentioned, ligands and receptors form a complex interaction network, where multiple ligands share receptors, potentially coupling their signaling. All these interactions are in turn coupled to *receptor trafficking*, which is known to be a mechanism that regulates signal transduction (Gonzalez-Gaitan, 2003; Lauffenburger and Linderman, 1993). Trafficking has been investigated in detail in many signal transduction pathways, such as the epidermal growth factor receptor (EGFR) and G protein-coupled receptor (GPCR) pathways (Bohm et al., 1997; Tan et al., 2004; Wiley et al., 2003). The typical way in which trafficking and signaling are coupled is by the induction of receptor internalization upon ligand binding and receptor activation, as for instance in the EGFR and GPCR pathways. After internalization, receptors can activate other signaling pathways, be modified in specific ways, and be targeted for degradation or recycling back to the plasma membrane.

A peculiarity of the TGF-β pathway is that receptors are constitutively internalized, even in the absence of ligand (Di Guglielmo et al., 2003; Mitchell et al., 2004). The trafficking route that the receptors follow, however, depends on whether or not they are in a signaling complex (Figure 3). Different routes will trigger different signaling outcomes and affect how receptors are degraded. Therefore, although the explicit implementation of the coupling is different than in the EGFR and GPCR pathways, receptor trafficking and signaling are also tightly coupled in the TGF-β pathway.

Taking all the current experimental data together, it is clear that many details of the underlying processes remain largely unknown. Recent experiments (Di Guglielmo et al., 2003; Mitchell et al., 2004), however, provide key information that allows computational modeling to bridge the gap between potential molecular mechanisms and experimentally observable behavior. The TGF-β pathway is at a discovery stage where modeling can help to gain "functional" intuition.

Here we characterize computationally the diverse potential types of behavior that the pathway structure itself can confer on the system. The types of behavior include responses to persistent changes in ligand concentration that can be transient or sustained and simultaneous responses to multiple ligands that can be passed downstream independently of or dependently on each other.

A sustained response implies that the steady-state signaling activity is a function of the ligand concentration. In this case, the higher the ligand concentration, the higher the activity of the pathway. For a transient response that precisely returns to the prestimulus level, in contrast, the steady-state activity is always the same and the pathway can only detect changes in ligand concentration. When multiple ligands signal in a dependent fashion, the extent of the coupling can be such that one ligand can suppress the effects of another one. In this regime, the pathway does not detect ligand concentrations but ratios of concentrations.

As we show here, all these types of behavior can be present in the TGF-β pathway. Which specific one is selected is determined by the interplay between trafficking and signaling. Thus the pathway can be set to detect, already at the receptor level, absolute levels of ligand, temporal changes in ligand concentration, and ratios of multiple ligands. Such a flexibility in the pathway behavior can lead to diverse physiological outcomes that have been associated with facilitated tumor progression.

## A concise computational model

In order to study the signal processing potential of the ligand-receptor network coupled with receptor trafficking, we assemble all the essential elements into a concise mathematical model that captures the logic of the underlying processes. The main goal is to represent as much complexity as possible by a small number of quantities that have direct experimental interpretation.

The essential elements we consider are (Figure 4):

i)      Ligands induce the formation of receptor complexes with type II and type I receptors.

ii)     Receptors and ligand-receptor complexes can be present in two spatially distinct compartments: plasma membrane and internalized endosomes.

iii)    The signaling activity is proportional to the number of ligand-receptor complexes that are present in the internalized endosomes.

iv)    Receptors and ligand-receptor complexes are continuously internalized into endosomes and recycled back to the plasma membrane (Di Guglielmo et al., 2003; Mitchell et al., 2004).

v)     Receptor degradation has a constitutive contribution, which is the same for free receptors and ligand-receptor complexes (Di Guglielmo et al., 2003; Mitchell et al., 2004).

vi)    Receptor degradation has a ligand-induced contribution, which affects only receptors that have been complexed with ligands (Di Guglielmo et al., 2003; Mitchell et al., 2004).

We use these elements to develop a mathematical model based on rate equations that describe the dynamics of both how different molecular species transform into each other and how they traffic between the different cellular compartments. We assume that internalization, recycling, and degradation rates are proportional to the number of receptors or ligand-receptor complexes; and that the formation of ligand-receptor complexes is proportional to the ligand concentration and to the type I and type II receptor concentrations.

In a first step towards characterizing the effects of the coupling of signaling with receptor trafficking, we consider that only a single type of ligand is present. Explicitly, we study how the components of the canonical TGF-β pathway —one ligand (TGF-β) and two receptors (Alk5 and TGFβRII), as emphasized in Figure 2— respond to changes in ligand concentration. The mathematical equations are shown in Figure 4. For typical trafficking rates (see Appendix I), this model closely reproduces the typical time courses of Smad phosphorylation upon addition of ligand (Figures 5a and 5b).

The computational model can be used also to analyze how different parameters affect the behavior of the system. For instance, the time during which the signaling activity rises is related to the time required for internalization and recycling of the receptors. Thus, the signal will peak, or stop rising, at about 30-60 min after ligand addition. If the internalization and recycling rates are changed, the position of the peak changes accordingly (Figures 5c and 5d).

Other transmembrane receptor pathways, such as the EGFR pathway, have much faster kinetics; the EGFR pathway reaches peak activity as fast as 5 min after stimulation (Yarden and Sliwkowski, 2001). The main reason for these differences is that most of the EGF receptors are present in the plasma membrane and they are ready to signal upon the addition of ligand. A similar kinetics is also observed for many GPCRs. In the TGF-β pathway internalization occurs continuously and only about 5-10% of the receptors are present in the plasma membrane at a given time (Di Guglielmo et al., 2003). The remaining 90-95% of the receptors are internalized in endosomes. Receptors need to be recycled from the endosomes back to the plasma membrane in order to be able to interact with the ligand; and this process takes about 30 min on average.

We can use the computational model to study the effects of different mechanisms, such as different forms of receptor degradation, on the behavior of the system. It has been observed that the addition of ligand can stimulate the degradation of the receptors in two different ways. On the one hand, receptors complexed with ligand in the plasma membrane can be internalized through a lipid raft–caveolar degradation pathway without becoming active signalers (Di Guglielmo et al., 2003). The behavior obtained in this case is the one we have already described (Figure 5a). On the other hand, ligand-receptor complexes can follow the standard internalization clathrin pathway, signal, and then be targeted for degradation upon returning to the plasma membrane (Mitchell et al., 2004). Interestingly, when this mechanism is explicitly modeled, the behavior obtained (Figure 5e) is qualitatively the same as the previous one (Figure 5a). Likewise, when both degradation mechanisms are considered together, a similar type of behavior is also obtained (Figure 5f). Thus, at this level of detail and for this range of parameter values, different mechanisms that implement ligand-induced degradation can lead to similar behavior.

## Control of the signal: transient vs. permanent responses

How is it possible to modify the form in which the system responds to changes in TGF-β concentration? A mathematical analysis of the model (see below and Theory Box for details) indicates that the key quantity that determines the qualitative behavior of the pathway is the ratio of the constitutive to the ligand-induced rate of degradation, referred to, in short, as the constitutive-to-induced degradation ratio (CIR). This quantity compares the rates of two degradation processes and, in general, does not have a simple expression in terms of rate constants.

Depending on the CIR, a permanent change in ligand concentration (as those in the insets of Figures 6a and 6b ) can elicit responses between two extremes:

  i) For low CIR, the ligand-induced degradation process dominates and there is a transient increase in signaling activity that returns to pre-stimulus levels (Figures 6a and 6b ).

  ii) For high CIR, the constitutive degradation process dominates and there is a permanently elevated level of signaling activity (Figures 6e and 6f).

For intermediate CIR, the behavior of the system is a mixture of both limiting types of behavior, with transient and permanent components (Figures 6c and 6d). The precise parameter values influence the amplitude and characteristic time of the response (see for instance Figures 5c and 5d), but its qualitative shape, that is, whether the response is transient (Figures 6a and 6b) or permanent (Figures 6e and 6f), depends only on the CIR.

The intuitive explanation of such types of behavior is as follows (for a detailed mathematical analysis see Appendix II). The probability for a receptor to bind the ligand, and therefore to become active, increases with the ligand concentration. If the ligand does not induce the degradation of receptors, the number of receptors remains constant and the total activity increases when the ligand concentration increases. If the ligand induces the degradation of the receptors, the number of receptors starts to decrease after ligand addition, which will eventually attenuate the signal. At steady state, the production and degradation of receptors equal each other. In the limit of the CIR going to zero, the signal adapts completely because degradation is proportional to the activation of receptors, and therefore activation is also proportional to the production of receptors. Thus, it is the receptor production rate, not the ligand concentration, that determines the steady-state signaling activity.

There are clear examples in other signal transduction pathways that show that these two limiting types of behavior can potentially lead to different physiological outcomes. For instance, transient activation of the MAPK cascade by EGF leads to cell proliferation. In contrast, permanent activation of the MAPK cascade by NGF leads to cell differentiation. In both cases, activation of the MAPK cascade induces the expression of a negative regulator that shuts down the activity of this cascade. The differences between EGF and NGF have been attributed to additional pathways activated by NGF that can prevent the inactivation of the MAPK cascade (Vaudry et al., 2002). Our model shows that such transient and permanent types of behavior can also be achieved by just changing the trafficking patterns, in particular by adjusting the CIR, without the need of explicitly expressing a negative regulator to shut down the cascade after signaling.

Remarkably, the duration of the signaling activity also seems to affect the physiological outcomes triggered by TGF-β (Nicolas and Hill, 2003). Epithelial cells that are sensitive to the antiproliferative effects of TGF-β (HaCaT and Colo-357) have sustained activity of more than 6 hours. In contrast, pancreatic tumor cell lines (PT45 and Panc-1) show short transient activity of about 1-2 hours. Such a short transient confers resistance to the antiproliferative effects of TGF-β but maintains other responses to TGF-β that can lead to increased malignancy and invasiveness (Nicolas and Hill, 2003). In our model, those differentiated types of behavior arise naturally for different trafficking patters. In particular, short transients and sustained responses imply a low and a high CIR, respectively.

## Control of the signals: coupled vs. uncoupled channels

*In vivo* conditions, in contrast to those typical of *in vitro* experiments, expose cells to complex environments with many different growth factors. When multiple ligands of the TGF-β superfamily are present at the same time, they are likely to affect each other's signaling (Figure 2). To study how multiple simultaneous input signals are integrated into coordinated transcriptional responses, we extend our computational model to consider two ligands that signal through two different type II receptors and a shared common type I receptor (see Appendix III for the mathematical equations). This is the simplest case of signal integration.

Intuitively, one should expect signals to be coupled when the shared receptor is saturated with ligands and uncoupled when ligand concentrations are low. At saturation, increasing the concentration of one ligand, and thus the concentration of the corresponding ligand-receptor complex, will take the shared receptor away from the complex formed by the other ligand, thus decreasing its signaling.

A mathematical analysis of the model (see Theory Box) indicates that even when the receptors are far from ligand-saturating conditions it is possible for signals to affect each other. The key element is again receptor trafficking. In essence, the coupling arises because the induction of degradation of the common receptor by one ligand attenuates the effects of the other ligand, which also requires the common receptor to transduce the signal.

For pathways working away from receptor saturation, the interplay between trafficking and signaling determines how multiple simultaneous signals are passed downstream. As in the single ligand case, there are two extreme types of behavior:
  (i)   For low CIR, the ligand-induced degradation process dominates and signals are completely coupled (Figures 7a and 7b).
  (ii)  For high CIR, the constitutive degradation process dominates and signaling is uncoupled (Figure 7e and 7f).

When the signals are completely coupled, the steady-state number of all ligand-receptor complexes remains constant and is independent of the ligand concentration. In this case, increasing one signal will decrease the other one by the same amount. When signals are uncoupled, the numbers of each species of ligand-receptor complexes change independently of each other. In general, for intermediate CIR, signals will show some degree of coupling (Figures 7c and 7d). These results demonstrate that changes in trafficking patterns, and the corresponding degradation, can alter the way in which the pathway integrates multiple, simultaneous signals.

The completely coupled case is especially interesting because it indicates that one ligand can potentially inhibit the effects of another one. Ligand-induced degradation is thus not only a mechanism for achieving transient responses, but also for coupling multiple signals. The fact that TGF-β can signal not only via Alk5 but also via Alk1 and Alk2 (Figure 2) potentially couples TGF-β signals to those of Activin A, BMP 6, BMP 7, and MIS. Thus, if TGF-β loses its growth suppressor properties, it could promote growth by inhibiting other growth suppressor pathways. For instance, there are dominant negative TGFβRII mutants that when overexpressed attenuate the response to TGF-β (Dumont and Arteaga, 2003). The presence of any of these mutant receptors and TGF-β results in the formation of futile receptor complexes that can target receptors for degradation, or take receptors away, that otherwise would be available to transduce the signals of other members of the TGF-β superfamily.

## Context-dependent response to TGF-β

The role reversal of TGF-β from negative to positive growth regulator is a widespread feature of tumor progression and is often associated with endogenous overexpression of TGF-β. As we have mentioned in the introduction, it is associated in some situations with the epithelial-to-mesenchymal transition (Gotzmann et al., 2004; Thiery, 2003). Under these conditions, the transcriptional program of tumor cells changes so that the Smad-activated genes promote rather than repress growth.

In other situations such a transition does not seem to be present. It has been observed in breast, prostate, and colon cancer cell lines that the action of TGF-β as growth promoter or suppressor depends on whether the cells were grown in an *in vitro* environment or *in vivo* in mouse xenografts (Steiner and Barrack, 1992; Tobin et al., 2002; Ye et al., 1999). The reasons for such a change remain largely unknown. It has been speculated that it could be a consequence of the effects of TGF-β on the *in vivo* microenvironment of the tumor cells. Another possibility is that other grow factors, such as EGF, affect how TGF-β is ultimately coupled to the cell cycle. Our model explicitly shows that the role reversal is a potentially intrinsic consequence of the design of the ligand-receptor interaction network and trafficking machineries and that can be the result of TGF-β attenuating the effects of growth-suppressing signals of other members of the TGF-β superfamily that might be present in the *in vivo* cell environment.

## Simultaneous perfect adaptation and coupled signaling

Our model also indicates that the conditions that give rise to completely coupled integration of multiple signals are the same that, in a single-ligand system, cause the signaling activity to completely adapt to its prestimulus level. Remarkably, this concurs with observations in prostate cancer cell lines, which show that the *in vivo* context can not only make TGF-β a growth promoter but also that the *in vitro* response to TGF-β is transient (Steiner and Barrack, 1992). One should expect the extracellular environment of growth factors to be more complex *in vivo* than *in vitro*. This relationship between *in vivo* and *in vitro* behavior and its connection to the coupling between receptor trafficking and signaling underscores the importance of understanding how signal transduction pathways are embedded within the cellular microenvironment under physiologically relevant conditions. Not only mutations in the canonical pathway but also changes in trafficking patterns can move the pathway to a different functioning point.

The qualitative results of our model, such as the regimes leading to transient and permanent responses as well as to completely coupled and uncoupled modes of signal integration, do not depend on the details of the model but on general properties (see Theory Box). Thus the main ideas are also relevant to other signal transduction pathways that are coupled to receptor trafficking. In particular, revisiting the experimental data, one can see that the interplay between adaptation and signal integration (Theory Box) is also present in the EGFR pathway (see figure 4 and 5 of reference (Worthylake et al., 1999)), in which down-regulation of erbB-2 by EGF concurs with adaptation of the signal transmitted by EGFR.

## Discussion

Cellular functions are controlled by networks of interacting molecules that operate at different levels of organization (Hartwell et al., 1999; Stelling et al., 2004; Vilar et al., 2003). Here, we have developed a concise computational model of the TGF-β pathway that shows that the receptors for the TGF-β superfamily of ligands are not just passive signal transducers. They are organized in a network that is able to process the signals before passing them downstream. Changes in receptor trafficking patterns can modify the type of behavior of the pathway in response to single and multiple ligand inputs. Already at the receptor level, the pathway can detect absolute levels of ligands, temporal changes in ligand concentration, and ratios of multiple ligands. This extra level of regulation can explain a wide variety of phenomena, such as the counterintuitive role reversal of TGF-β from suppressor to promoter of growth, and leads to a unified interpretation of seemingly disparate experimental observations. A key quantity that determines the qualitative behavior of the pathway is the constitutive-to-induced degradation ratio (CIR) of the receptors. For low CIR, the pathway responds transiently to sustained changes in ligand concentration and the signaling activities of multiple simultaneous ligands become dependent on each other. Ligand-induced degradation is thus not only a mechanism for achieving transient responses and perfect adaptation, but also for coupling multiple signals.

## Acknowledgments

We are indebted to Joan Massagué and Van Le for invaluable help during the early stages of this work. We are also very grateful to Gary Bader, Debora Marks, Leonor Saiz, Nikolaus Schultz, Wenying Shou, and Stas Shvartsman for numerous discussions, comments, and suggestions. C.S. research was funded in part by the Alfred W. Bressler Scholars Endowment Fund.

## Theory Box: Coupled signaling and perfect adaptation

Computational modeling offers precise insights into the functioning of the TGF-β pathway. It is possible to go a step further and generalize the conditions that give rise to different qualitative types of behavior.

Let us consider two ligands ($l_1$ and $l_2$), one type I receptor ($R_I$) and two type II receptors ($R_{II,1}$ and $R_{II,2}$). The type I receptor is shared among the two ligand-receptor complexes [$l_1\,R_I\,R_{II,1}$] and [$l_2\,R_I\,R_{II,2}$]. The following conservation equations refer to the common type I receptor at steady-state.

Under stationary conditions, the number of receptors produced (by gene expression) is equal to the number of receptors degraded:

$$p = d_{const} + d_{lid} \,,$$

where $p$ is the receptor production rate; and $d_{const}$ and $d_{lid}$ are the constitutive and ligand-induced degradation rates, respectively. Assuming that a fraction $\delta$ of the activated receptors is degraded through a ligand-induced degradation process, we can express $d_{lid}$ as

$$d_{lid} = \delta(i_{a1} + i_{a2}) \text{ with } \delta \in (0, 1) \,,$$

where $i_{a1}$ and $i_{a2}$ are the rates of formation of the ligand-receptor complexes ([$l_1\,R_I\,R_{II,1}$] and [$l_2\,R_I\,R_{II,2}$], respectively). Therefore,

$$p = d_{const} + \delta(i_{a1} + i_{a2}) \,.$$

We now explicitly consider the following two limiting cases:

## Case 1: no ligand-induced degradation ($d_{const} > 0$, $\delta = 0$)

In this case, $d_{lid} = 0$, which leads to $p = d_{const}$. Since $d_{const} \equiv d_{const}(R_T)$ is a function of the total number of receptors $R_T$, the previous condition indicates that the number of receptors remains constant $R_T = d_{const}^{-1}(p)$, where $d_{const}^{-1}$ is the inverse function of $d_{const}$. For instance, if the constitutive receptor degradation follows first order kinetics, $d_{const} = \gamma R_T$, then $R_T = p/\gamma$. Under these conditions, if the rate of formation of complexes ($i_{a1} + i_{a2}$) is small (for instance, for low ligand concentrations) compared with constitutive internalization and degradation, there is no coupling between signaling channels.

## Case 2: only ligand-induced degradation ($d_{const} = 0$, $\delta > 0$)

In this case $d_{const} = 0$ and $p = \delta(i_{a1} + i_{a2})$. This implies that the formation of one ligand-receptor complex excludes the formation of the other one. In this case, the number of receptors in the plasma membrane does not remain constant, but is adjusted so that for a given ligand concentration the rate of formation of complexes ($i_{a1} + i_{a2}$) remains equal to $p/\delta$. As an explicit example, the kinetics $i_{a1} = l_1\,R_I\,R_{II,1}$ and $i_{a2} = l_1\,R_I\,R_{II,2}$ implies $R_I =$ $p/(l_1\ R_{II,1} + l_2\ R_{II,2})$.   Under these conditions, the completely coupled mode of signal integration arises even for low ligand concentrations.

## Connection with the temporal behavior

The fact that the rate of formation of complexes remains constant implies that for the single ligand case (for example, $i_{a2} = 0$) we have $p = \delta i_{a1}$ when there is only ligand-induced degradation.  The steady state of the system is fixed no matter what the ligand concentration is; consequently, changes in ligand concentration can only elicit transient responses that completely adapt to the prestimulus level.  The system exhibits perfect adaptation (Csete and Doyle, 2002).

## Appendix I: Trafficking rates

*Internalization rate*. It has been reported in Table 1 of reference (Di Guglielmo et al., 2003) that after 15 min of labeling the receptors at the plasma membrane only 8, 6, 4, and 2 % of the labeled receptors remain at the plasma membrane (the different percentage values correspond to different experimental conditions). The remaining labeled receptors have been internalized in either caveolin-1 positive or caveolin-1 negative vesicles. By using the formula $k_i = -\dfrac{1}{t}\ln f_t$, where $k_i$ is the internalization rate and $f_t$ the fraction of labeled receptors that remain at the plasma membrane after a time $t$, we obtain internalization rates of 1/5.9, 1/5.3, 1/4.7, and 1/3.8 $\min^{-1}$, respectively. We have chosen $k_i = 1/(3\,\text{min})$ for comparison with experimental data in figure 5b.

Figure 4 of reference (Mitchell et al., 2004) shows that 1.7% of the total number of receptors are internalized per minute. When this value is rescaled by the fraction of receptors in the plasma membrane, it translates into 18% of surface receptors internalized per minute. This implies that this internalization rate is $k_i = 1/(5.3\,\text{min})$, which is similar to the results obtained from reference (Di Guglielmo et al., 2003).

*Ligand-induced degradation rate*. Active ligand-receptor complexes in lipid raft–caveolar compartments can recruit Smad7-Smurf2 (Monteleone et al., 2004), which then targets them for degradation (Di Guglielmo et al., 2003). Reference (Di Guglielmo et al., 2003) shows that receptors are internalized through the clathrin pathway and lipid raft–caveolar compartments with similar rates. We have chosen $k_{lid} = 1/(4\,\text{min})$ for comparison with experimental data in figure 5b.

*Constitutive degradation rate*. Figure 3 of reference (Di Guglielmo et al., 2003) shows that when the lipid raft–caveolar pathway is blocked with nystatin, only ~30% of the initially labeled receptors remain in the cell after 8 hours. This gives a characteristic degradation time of ~400 min with respect to the total number of receptors. Rescaling this number to the plasma membrane receptors we obtain $k_{cd} = -\dfrac{k_r}{k_r + k_i}\dfrac{1}{480\,\text{min}}\ln 0.3 = 1/(36\ \text{min})$. We have chosen $k_{cd} = 1/(36\,\text{min})$ for comparison with experimental data in figure 5b.

*Recycling rate*. Figure 3 of reference (Mitchell et al., 2004) shows that after about 30 min cells stop secreting internally labeled TGF-β receptors. This recycling rate is similar to that for the EGF receptor. We have chosen $k_r = 30\,\text{min}$ for comparison with experimental data in figure 5b.

## Appendix II: Steady and quasi-steady state analysis

In this appendix we study the steady state of the system with a single ligand. By equating to zero the derivatives in the model equations of Figure 4, we obtain that the steady-state number of internalized ligand-receptor complexes is

$$\overline{[lR_I R_{II}]} = \frac{k_i}{k_r} \frac{k_a[l][R_I][R_{II}]}{(k_{cd} + k_{lid} + k_i)},$$

where the steady-state number of type I and type II receptors at the plasma membrane are obtained by solving the equations

$$0 = a_I - c[R_I][R_{II}] - [R_I],$$

$$0 = a_{II} - c[R_I][R_{II}] - [R_{II}],$$

with

$$a_I = \frac{p_{RI}}{k_{cd}}, \quad a_{II} = \frac{p_{RII}}{k_{cd}}, \quad \text{and} \quad c = \frac{k_{cd} + k_{lid} + (1-\alpha)k_i}{(k_{cd} + k_{lid} + k_i)} \frac{k_a}{k_{cd}}[l].$$

The solution of these equations is

$$[R_I] = \frac{a_I - a_{II}}{2} + \frac{\sqrt{1 + 2(a_I + a_{II})c + (a_I - a_{II})^2 c^2} - 1}{2c},$$

$$[R_{II}] = \frac{a_{II} - a_I}{2} + \frac{\sqrt{1 + 2(a_{II} + a_I)c + (a_{II} - a_I)^2 c^2} - 1}{2c},$$

which leads to

$$\overline{[lR_I R_{II}]} = \frac{k_i k_{cd}(1 + (a_I + a_{II})c - \sqrt{1 + 2(a_I + a_{II})c + (a_I - a_{II})^2 c^2})}{k_r(k_{cd} + k_{lid} + (1-\alpha)k_i)2c}.$$

For low values of $c$, the previous equation reduces to

$$\overline{[lR_I R_{II}]} = \frac{k_i k_{cd} a_I a_{II} c}{k_r(k_{cd} + k_{lid} + (1-\alpha)k_i)},$$

which indicates that the steady-state number of internalized ligand-receptor complexes is proportional to the ligand concentration and the production of each receptor type.

For high values of $c$, in contrast, we obtain

$$\overline{[lR_I R_{II}]} = \frac{k_i k_{cd} a_{II}}{k_r(k_{cd} + k_{lid} + (1-\alpha)k_i)} \quad \text{if} \quad a_{II} \leq a_I,$$

and

$$\overline{[lR_I R_{II}]} = \frac{k_i k_{cd} a_I}{k_r(k_{cd} + k_{lid} + (1-\alpha)k_i)} \quad \text{if} \quad a_I \leq a_{II}.$$

Therefore, for high ligand concentration or low constitutive degradation the steady-state number of internalized ligand-receptor complexes is controlled by the receptor with the smallest production rate and this number does not depend on the ligand concentration.

The case of high $c$ and low constitutive degradation is specially interesting because the steady-state signal does not depend on the ligand concentration, even when the ligand is present in small quantities. An important question to address now is: can the system detect changes in concentration in this regime? When the recycling rate is much lower than the internalization rate, the number on ligand receptor complexes in the plasma membrane equilibrates faster than all the other variables. Therefore assuming quasi-equilibrium in this variable, $\frac{d}{dt}[lR_IR_{II}] = 0$, we obtain that, upon changes in the ligand concentration ($\Delta[l]$), the changes in the number of ligand receptor-complexes in the plasma membrane ($\Delta[lR_IR_{II}]$) follow the equation

$$0 = k_a([l]+\Delta[l])([R_I]-\Delta[lR_IR_{II}])([R_{II}]-\Delta[lR_IR_{II}]) - (k_{cd}+k_{lid}+k_i)([lR_IR_{II}]+\Delta[lR_IR_{II}]).$$

Note that we have assumed that the number of receptors in the plasma membrane is conserved at these time scales. For small changes in ligand concentration, we obtain

$$\Delta[lR_IR_{II}] = \frac{k_a\Delta[l][R_I][R_{II}]}{k_a[l]([R_I]+[R_{II}])+k_{cd}+k_{lid}+k_i}.$$

This expression indicates that for high $c$, low constitutive degradation, and slow recycling (compared to internalization) the system can detect changes in ligand concentration while keeping a steady-state signal that does not depend on ligand concentration.

## Appendix III: Two-compartment model of receptor trafficking for two ligands

The equations for a system with two ligands with concentrations $[l_1]$ and $[l_2]$ are:

$$\frac{d}{dt}[l_1 R_I R_{II,1}] = k_a[l_1][R_I][R_{II,1}] - (k_{cd} + k_{lid} + k_i)[l_1 R_I R_{II,1}]$$

$$\frac{d}{dt}[l_2 R_I R_{II,2}] = k_a[l_2][R_I][R_{II,2}] - (k_{cd} + k_{lid} + k_i)[l_1 R_I R_{II,2}]$$

$$\frac{d}{dt}[R_I] = p_{RI} - k_a[l_1][R_I][R_{II,1}] - k_a[l_2][R_I][R_{II,2}] - (k_{cd} + k_i)[R_I] + k_r[\overline{R_I}]$$
$$+ \alpha k_r[\overline{l_1 R_I R_{II,1}}] + \alpha k_r[\overline{l_2 R_I R_{II,2}}]$$

$$\frac{d}{dt}[R_{II,1}] = p_{RII,1} - k_a[l_1][R_I][R_{II,1}] - (k_{cd} + k_i)[R_{II,1}] + k_r[\overline{R_{II,1}}]$$
$$+ \alpha k_r[\overline{l_1 R_I R_{II,1}}]$$

$$\frac{d}{dt}[R_{II,2}] = p_{RII,2} - k_a[l_2][R_I][R_{II,2}] - (k_{cd} + k_i)[R_{II,2}] + k_r[\overline{R_{II,2}}]$$
$$+ \alpha k_r[\overline{l_2 R_I R_{II,2}}]$$

$$\frac{d}{dt}[\overline{l_1 R_I R_{II,1}}] = k_i[l_1 R_I R_{II,1}] - k_r[\overline{l_1 R_I R_{II,1}}]$$

$$\frac{d}{dt}[\overline{l_2 R_I R_{II,2}}] = k_i[l_1 R_I R_{II,2}] - k_r[\overline{l_2 R_I R_{II,2}}]$$

$$\frac{d}{dt}[\overline{R_I}] = k_i[R_I] - k_r[\overline{R_I}]$$

$$\frac{d}{dt}[\overline{R_{II,1}}] = k_i[R_{II,1}] - k_r[\overline{R_{II,1}}]$$

$$\frac{d}{dt}[\overline{R_{II,2}}] = k_i[R_{II,2}] - k_r[\overline{R_{II,2}}]$$

The variables $[R_I]$, $[R_{II,1}]$, and $[R_{II,2}]$ are the numbers of type I and type II receptors in the plasma membrane; and $[l_1 R_I R_{II,1}]$ and $[l_2 R_I R_{II,2}]$ refer to the corresponding ligand-receptor complexes. The overline indicates internalized receptors and ligand-receptor complexes. The signaling activity triggered by each ligand is assumed to be proportional to the corresponding number of internalized ligand-receptor complexes. $k_a$ is the rate constant of ligand-receptor complex formation; $p_{RI}$, $p_{RII,1}$ and $p_{RII,2}$ are the rates of receptor production; $k_i$, $k_r$, $k_{cd}$, and $k_{lid}$ are the internalization, recycling, constitutive degradation, and ligand-induced degradation rate constants; and $\alpha$ is the fraction of active receptors that are recycled back to the plasma membrane and can interact again with the ligand.

## *Figure Legends*

## Figure 1

## Formation of receptor hetero-tetramers

The active form of the TGF-β ligand is a dimer of two molecules held together by hydrophobic interactions and a disulfide bond (Massague, 1998; Sun and Davies, 1995). This dimer induces the formation, at the plasma membrane, of receptor hetero-tetramers that contain two type I and two type II receptors (Goumans et al., 2003; Yamashita et al., 1994). The type II receptors phosphorylate the type I receptors; the type I receptors are then enabled to phosphorylate cytoplasmic R-Smads, which then act as transcriptional regulators.

**Figure 2**

**Interactions among the ligands of the TGF-β superfamily and their receptors**

(based on data reviewed in reference (de Caestecker, 2004))

The graphical representation lays out the specific type II/type I receptor complexes that different ligands mediate. Each set of links drawn between a type II and type I receptor, mediated by a connecting ligand, represents a feasible ligand-receptor complex. The 14 ligands, 5 type II and 7 type I receptors shown here give rise to 50 different combinations of ligand-receptor complexes overall. Note that many of these 50 complexes share ligand and receptor species.

The ligand-receptor complexes phosporylate the cytoplasmic R-Smads; at this point the signal is essentially funneled into two different pathways. The decision which one is chosen depends on the particular type I receptor in the ligand-receptor complex. The type I receptors can be divided into two groups, depending on which subgroup of R-Smads they bind and phosphorylate: The first group of type I receptors (Alk1/2/3/6, shown on the top right) bind and activate the R-Smads Smad1/5/8, whereas the second group (Alk4/5/7, shown on the bottom right) act on the R-Smads Smad2/3. The phosphorylated R-Smads then form complexes with the Co-Smad Smad4.

**Figure 3**

**Signaling and trafficking in the TGF-β pathway**

Receptors in the plasma membrane interact with the signaling peptides of the TGF-β superfamily to form active complexes. Receptors and activated ligand-receptor complexes can internalize via clathrin-coated pits into endosomes, from where the active ligand-receptor complexes phosphorylate the cytoplasmic R-Smads ("receptor Smads", either the Smad1/5/8 or the Smad2/3 group) (Shi and Massague, 2003).

The phosphorylated R-Smads form complexes with the Co-Smad (Smad4) and then translocate into the nucleus where they act as transcriptional regulators of about 300 target genes.

The internalized receptors recycle back to the plasma membrane (with a characteristic time of ~30 minutes) via a rab11-dependent, rab4-independent pathway (Mitchell et al., 2004). After returning to the plasma membrane, the receptors that were actively signaling can be targeted for degradation or be used for further ligand-binding or internalization (Mitchell et al., 2004). Receptors that did not bind ligands are simply returned to the plasma membrane. As a consequence of the trafficking processes only about 5-10% of receptors are present in the plasma membrane (Di Guglielmo et al., 2003).

In addition to the traditional clathrin pathway, active ligand receptor-complexes can recruit Smad7-Smurf2 (Monteleone et al., 2004), which then targets them to lipid raft–caveolar compartments (right) for degradation (Di Guglielmo et al., 2003).

The ligands do not return back to the plasma membrane, but disassociate from the receptors before recycling and undergo direct degradation via the lysosomes (Mitchell et al., 2004).

Note that, in addition to the ligand-induced receptor degradation, we also consider a receptor degradation pathway that functions independently of ligand-binding; this represents a "constitutive" or ligand-independent degradation pathway (left).

**Figure 4**

**Two-compartment model of receptor trafficking and signaling**

Graphical representation and equations for a model with one ligand that forms complexes with one type I and one type II receptor. Receptors are present in two main compartments: the plasma membrane (receptors at the cell surface) and the endosomes (internalized receptors). Receptors and ligand-receptor complexes traffic between these two compartments by internalization and recycling. Only internalized ligand-bound receptors have kinase activity. Active receptors can also be internalized in a degradation pathway (right). In addition, receptors in the plasma membrane can undergo constitutive degradation, independently of whether they are ligand-bound (left). A supply of new receptors is constantly produced by gene expression.

The concentration of the ligand is denoted by $[l]$; the numbers of type I and II receptor and ligand-receptor complexes in the plasma membrane, by $[R_I]$, $[R_{II}]$, and $[l\,R_I R_{II}]$, respectively; and the numbers of internalized type I and II receptor and ligand-receptor complexes by $[\overline{R_I}]$, $[\overline{R_{II}}]$, and $[\overline{l\,R_I R_{II}}]$, respectively. (Note that type II receptors are not shown in the graphical representation.) The signaling activity of the pathway is assumed to be proportional to the number of internalized ligand-receptor complexes, $[\overline{l\,R_I R_{II}}]$. $k_a$ is the rate constant of ligand-receptor complex formation; $p_{RI}$ and $p_{RII}$ are the rates of receptor production; $k_i$, $k_r$, $k_{cd}$, and $k_{lid}$ are the internalization, recycling, constitutive degradation, and ligand-induced degradation rate constants; $\alpha$ is the fraction of active receptors that are recycled back to the plasma membrane and can interact again with the ligand.

**Figure 5**

**Typical time courses of the number of active receptor complexes upon addition of TGF-β**

The typical response to sustained changes in TGF-β concentration shows partial adaptation after reaching a maximum of activity. Different values of the parameters of the model lead to this characteristic behavior. In all panels, the TGF-β concentration is increased at time zero to saturating values and kept constant afterwards, as in panel a) inset.

**a)** Behavior of the model for typical trafficking rates: internalization, $k_i = (3 \text{ min})^{-1}$; recycling, $k_r = (30 \text{ min})^{-1}$; constitutive degradation, $k_{cd} = (36 \text{ min})^{-1}$; ligand-induced degradation $k_{lid} = (4 \text{ min})^{-1}$; efficiency of recycling of active receptors, $\alpha = 1$. Note that the trafficking rate constants are given as the inverse of the corresponding characteristic times. The production of receptors is $p_{RI} = 8 \text{ min}^{-1}$ and $p_{RII} = 4 \text{ min}^{-1}$. The units of ligand concentration are chosen so that the association rate constant is the unity, $k_a = 1$. For these units, EC$_{50} \approx 0.0002$. At time zero the ligand concentration changes from 0.00003 to 0.01. The signal peaks at ~60 min.

**b)** Comparison of the model time course (upper lane) with an experimental time course of nuclear, phosphorylated Smad2 (P-Smad) as reported by Inman et al. (bottom lane) (Inman et al., 2002). The model results from panel a are shown at the experimental time points and color-coded to ease comparison.

**c)** Behavior of the model with the same parameter values as in panel a, with the exception of the rate constants for *internalization and recycling* that have been *decreased* to $k_i = (10 \text{ min})^{-1}$ and $k_r = (100 \text{ min})^{-1}$. The signal peaks at ~180 min.

**d)** Behavior of the model with the same parameter values as in panel a, with the exception of the rate constants for *internalization and recycling* that have been *increased* to $k_i = (1 \text{ min})^{-1}$ and $k_r = (10 \text{ min})^{-1}$. The signal peaks at ~20 min.

**e)** Behavior of the model with the same parameter values as in panel a, with the exception of the rate constant for *ligand-induced degradation* that has been *decreased* to $k_{lid} = 0$ and the *efficiency of recycling* of active receptors that has been *decreased* to $\alpha = 0.5$. This implies that ligand-receptor complexes are not degraded via the caveolae pathway. In contrast, 50% of the active ligand-receptors that come back to the plasma membrane after they have signaled are degraded.

**f)** Behavior of the model with the same parameter values as in panel a, with the exception of the efficiency of recycling of active receptors that has been decreased to $\alpha = 0.5$. These parameters account for both types (caveola-dependent and recycling-dependent) of ligand-induced degradation.

## Figure 6

### Control of the kinetic signaling behavior

A key control quantity of the qualitative behavior of the system is the constitutive-to-induced degradation ratio (CIR). Panels on the left (a, c, and e) show the typical behavior of the system for different CIR values. The TGF-β concentration is increased at time zero to saturating values and remains constant afterwards (panel a inset). Panels on the right (b, d, and f) show the behavior of the system for the same parameter values as the corresponding panels on the left but when TGF-β concentration is increased slowly (panel b inset).

**a,b)** Same parameter values as in Figure 5a with the exception that the *constitutive* and *ligand-induced* degradation rates have been *decreased* and *increased* by a factor three, respectively: $k_{cd} = (3 \times 36 \text{ min})^{-1}$; ligand-induced degradation $k_{lid} = (4/3 \text{ min})^{-1}$.

**c,d)** Same parameter values as in Figure 5a. Figure in panel c is exactly the same as Figure 5a.

**e,f)** Same parameter values as in Figure 5a with the exception that the *constitutive* and *ligand-induced* degradation rates have been *increased* and *decreased* by a factor three, respectively: $k_{cd} = (36/3 \text{ min})^{-1}$; ligand-induced degradation $k_{lid} = (3 \times 4 \text{ min})^{-1}$.

**Figure 7**

**Control of signal integration**

Time courses of the numbers of active receptor complexes when TGF-β concentration (in red) is increased, repeatedly in steps or continuously. Left panels (a, c, and e) show the responses to step increases as shown in panel a inset. Right panels (b, d, f) show the response to a continuous increase as shown in panel b inset. There is also a second ligand present (here BMP7, in blue) whose concentration is kept constant. The two ligands induce the formation of two ligand-receptor complexes, $C_{BMP7}$ (blue) and $C_{TGF-\beta}$ (red), that share the type I receptor Alk2. The green line on the left panels shows the total number of active receptor complexes ($C_{BMP7} + C_{TGF-\beta}$). As in Figure 6, a key control quantity of the qualitative behavior of the system is the constitutive-to-induced degradation ratio (CIR). The parameter values for panels a, b, c, d, e, and f are the same as in Figures 6a, 6b, 6c, 6d, 6e, and 6f, respectively.

**Figure 1**

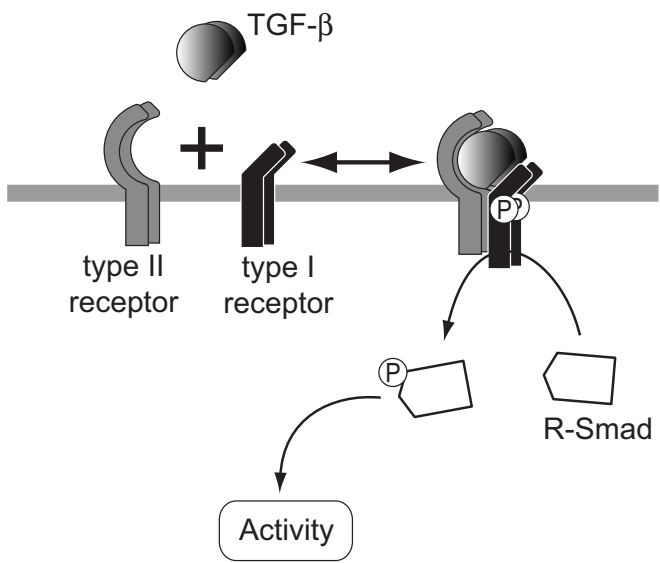

TGF-β

type II
receptor

type I
receptor

R-Smad

Activity

**Figure 2**

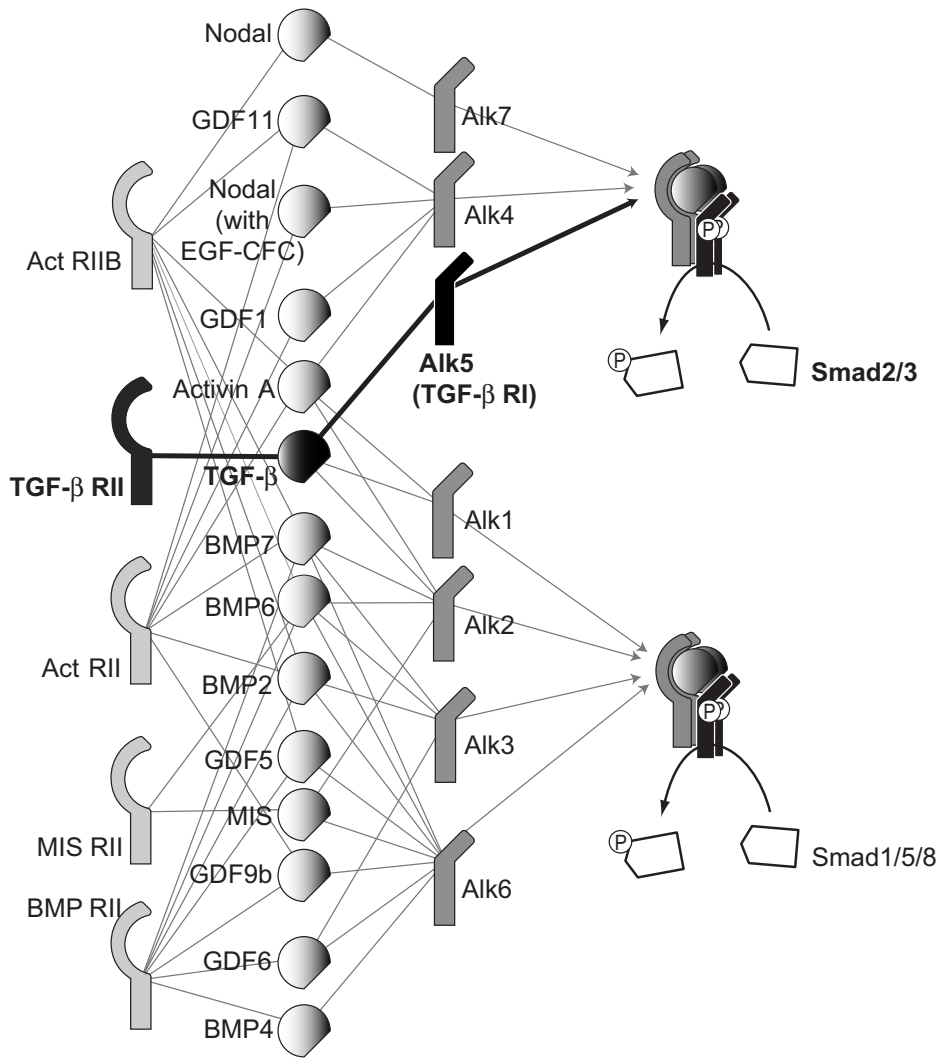

**Figure 3**

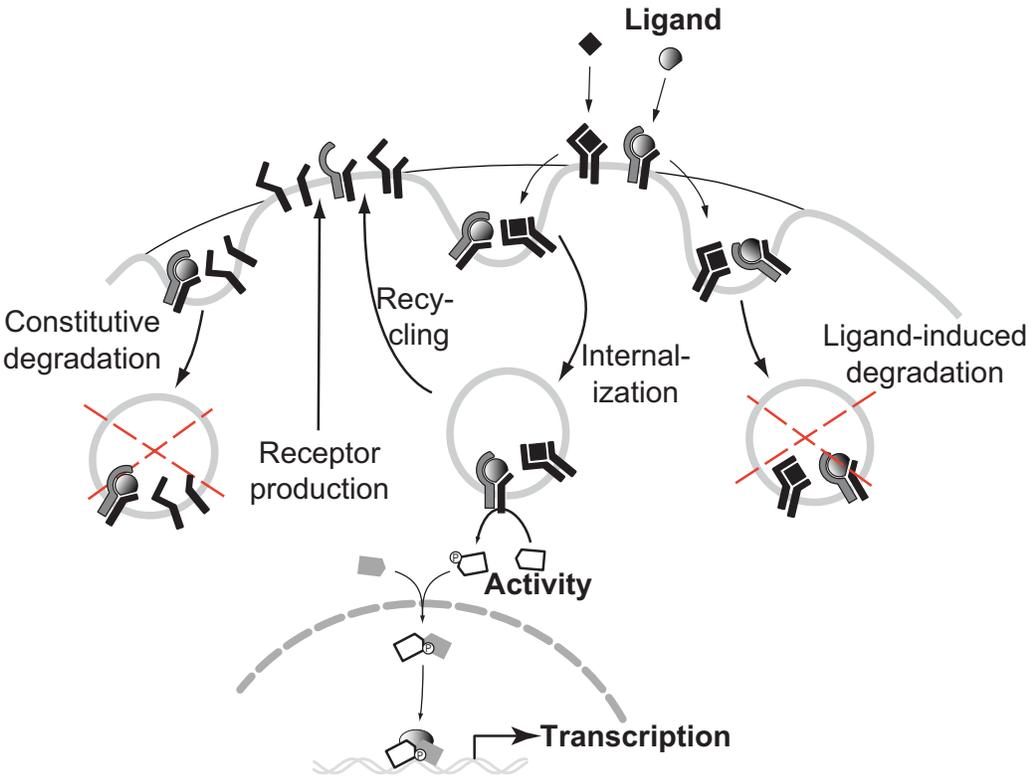

**Figure 4**

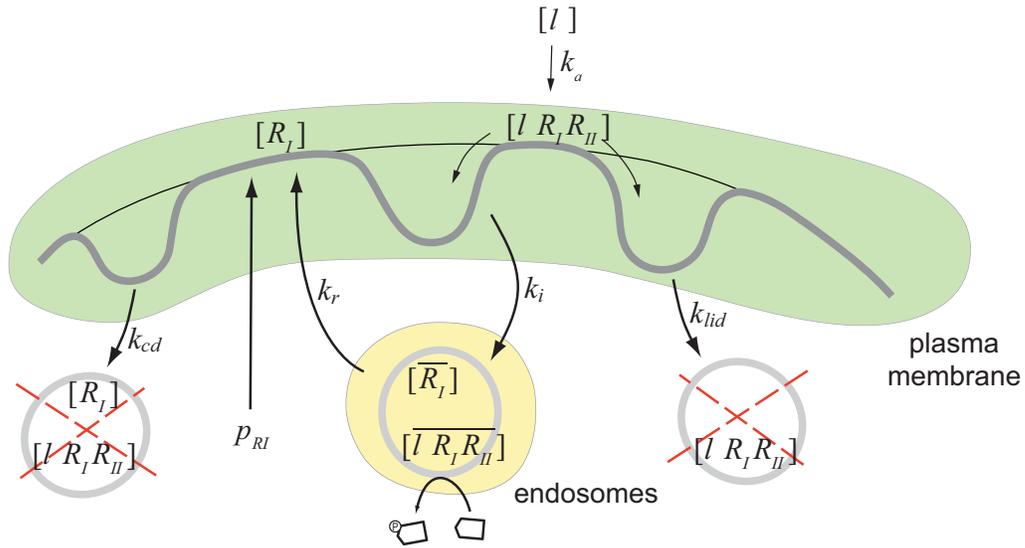

$$\frac{d}{dt}[l\ R_I R_{II}] = k_a[l\ ][R_I][R_{II}] - (k_{cd} + k_{lid} + k_i)[l\ R_I R_{II}]$$

$$\frac{d}{dt}[R_I] = p_{RI} - k_a[l\ ][R_I][R_{II}] - (k_{cd} + k_i)[R_I] + k_r[\overline{R_I}] + \alpha k_r[\overline{l\ R_I R_{II}}]$$

$$\frac{d}{dt}[R_{II}] = p_{RII} - k_a[l\ ][R_I][R_{II}] - (k_{cd} + k_i)[R_{II}] + k_r[\overline{R_{II}}] + \alpha k_r[\overline{l\ R_I R_{II}}]$$

$$\frac{d}{dt}[\overline{l\ R_I R_{II}}] = k_i[l\ R_I R_{II}] - k_r[\overline{l\ R_I R_{II}}]$$

$$\frac{d}{dt}[\overline{R_I}] = k_i[R_I] - k_r[\overline{R_I}]$$

$$\frac{d}{dt}[\overline{R_{II}}] = k_i[R_{II}] - k_r[\overline{R_{II}}]$$

**Figure 5**

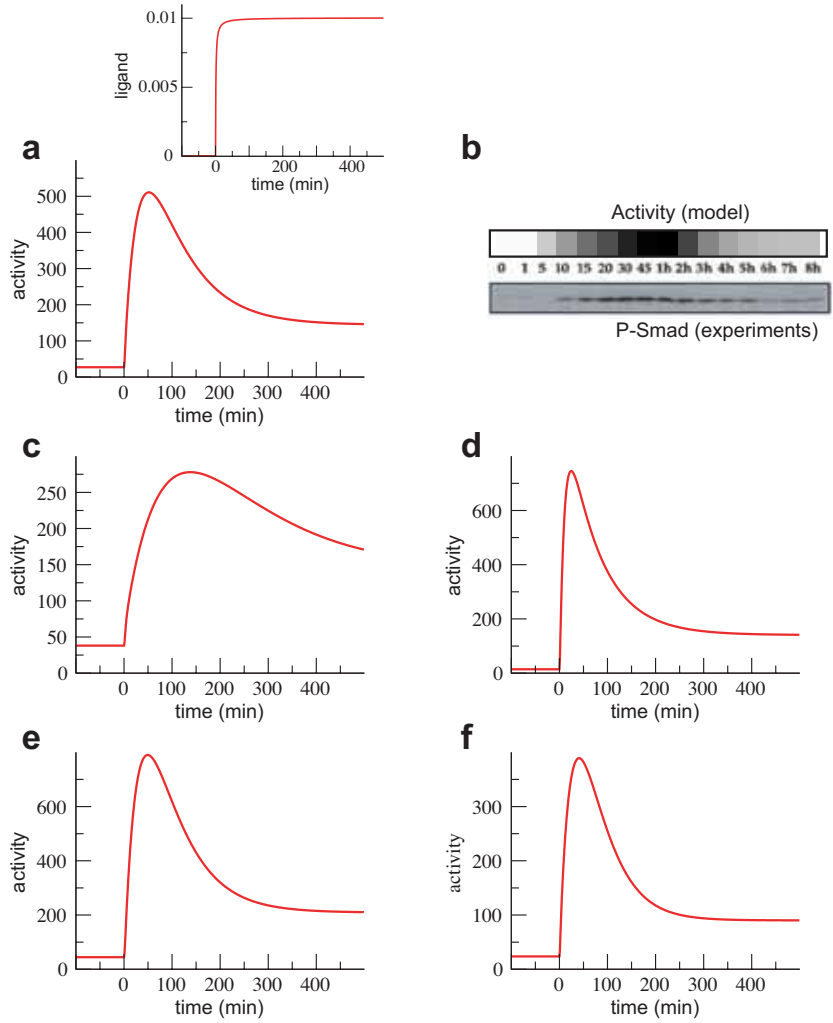

**Figure 6**

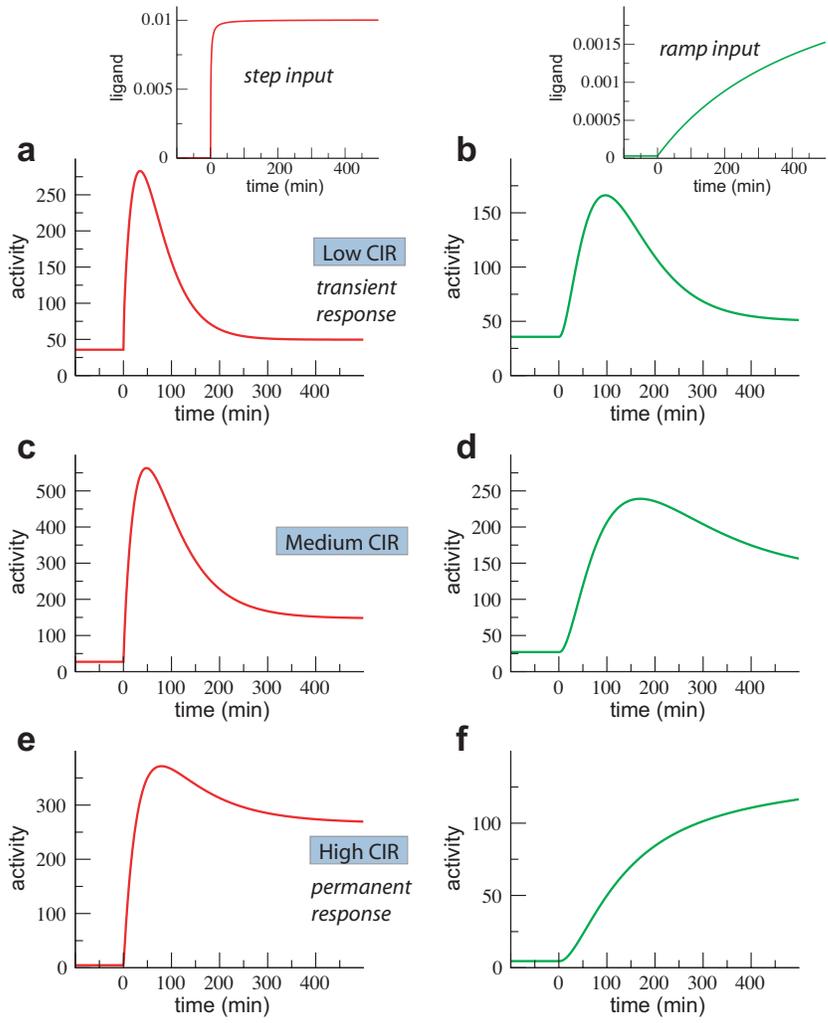

**Figure 7**

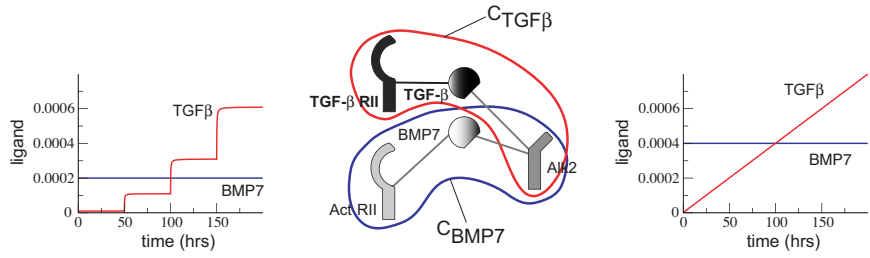

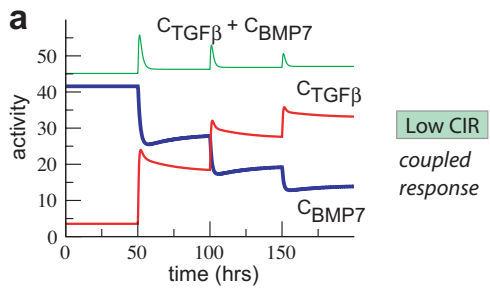

**a**

$C_{TGF\beta} + C_{BMP7}$

$C_{TGF\beta}$

$C_{BMP7}$

activity

time (hrs)

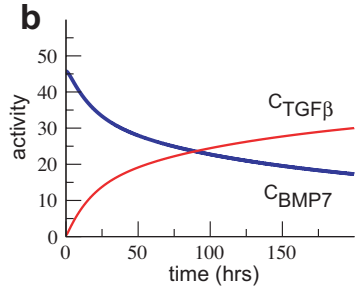

**b**

$C_{TGF\beta}$

$C_{BMP7}$

activity

time (hrs)

Low CIR

*coupled response*

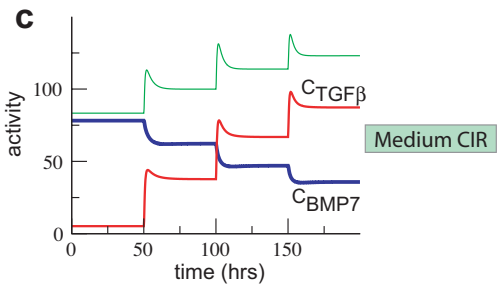

**c**

$C_{TGF\beta}$

$C_{BMP7}$

activity

time (hrs)

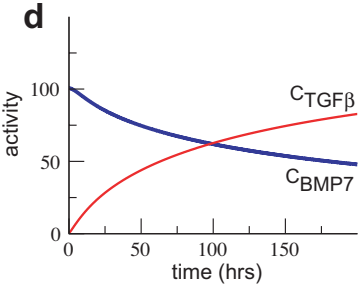

**d**

$C_{TGF\beta}$

$C_{BMP7}$

activity

time (hrs)

Medium CIR

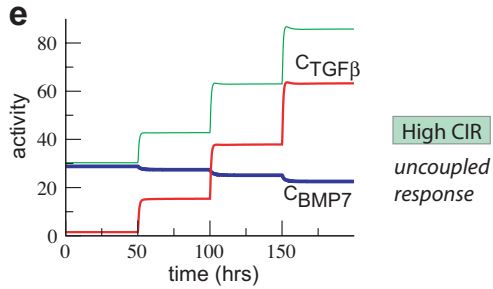

**e**

$C_{TGF\beta}$

$C_{BMP7}$

activity

time (hrs)

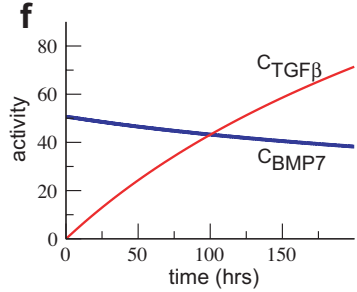

**f**

$C_{TGF\beta}$

$C_{BMP7}$

activity

time (hrs)

High CIR

*uncoupled response*